# DEEP LEARNING-GUIDED SURFACE CHARACTERIZATION FOR AUTONOMOUS HYDROGEN LITHOGRAPHY

*Mohammad Rashidi[‡1], Jeremiah Croshaw[‡*1], Kieran Mastel[1], Marcus Tamura[1], Hedieh Hosseinzadeh[3], and Robert A. Wolkow[*1,2,3]*

[1]Department of Physics, University of Alberta, Edmonton, Alberta, T6G 2J1

[2]Nanotechnology Research Centre, National Research Council Canada, Edmonton, Alberta, T6G 2M9, Canada

[3]Quantum Silicon, Inc., Edmonton, Alberta, T6G 2M9, Canada

**Abstract**

As the development of atom scale devices transitions from novel, proof-of-concept demonstrations to state-of-the-art commercial applications, automated assembly of such devices must be implemented. Here we present an automation method for the identification of defects prior to atomic fabrication via hydrogen lithography using deep learning. We trained a convolutional neural network to locate and differentiate between surface features of the technologically relevant hydrogen-terminated silicon surface imaged using a scanning tunneling microscope. Once the positions and types of surface features are determined, the predefined atomic structures are patterned in a defect-free area. By training the network to differentiate between common defects we are able to avoid charged defects as well as edges of the patterning terraces. Augmentation with previously developed autonomous tip shaping and patterning modules allows for atomic scale lithography with minimal user intervention.

**Introduction**

With the miniaturization of complementary metal-oxide-semiconductor (CMOS) technology approaching its fundamental limit, attention has been focused on developing alternatives built at the atomic level.[1,2,3] If these devices are to be commercially viable, they must be built in a way that allows parallelized and automated fabrication. Scanning Probe Microscopy (SPM) has provided a means for several different varieties of atom-scale device fabrication including memory systems using a Cu/Cl system[4] or dangling bonds (DBs) on hydrogen-terminated silicon (H-Si),[5] spin-based logic using Fe atoms on a Cu(111) surface,[6] single-atom transistors using phosphorus dopants in silicon,[7] and binary atomic wires and logic gates using DBs on the H-Si surface.[8] Despite the progress made in the design of these and other device concepts,[4,9,10,11] reliable device fabrication is usually limited by patterning accuracy or variability in the fabrication process. Dangling bonds on the H-Si surface have been shown to be rewritable[12,13,5] as well as stable at room temperature[14,15] making them an excellent candidate for atom scale devices.

The H-Si surface has found applications in the study of surface chemistry including self-directed growth of ordered multi-molecular lines[16,17] and reaction energetics.[18] The controllable desorption of hydrogen from the H-Si(100)-2x1 surface using the probe tip of a scanning tunneling microscope (STM),[19] allowed for more precise studies of surface chemistry[20] including fabrication of rudimentary devices.[21,22] With the continued study of DBs on H-Si surfaces, more complex and functional devices have been developed; however, complete automation of the hydrogen lithography process has been limited by three major factors. First, the probe requires continuous monitoring to ensure an atomically sharp patterning condition. This step was recently automated using machine learning.[23] Second, which is the subject of future work, is automated hydrogen lithography error detection and correction through recently realized controlled hydrogen repassivation techniques.[12,13] Third is the automated characterization and localization of defects on the H-Si(100)-2x1 surface to assign an ideal patterning area by avoiding certain charged and uncharged defects which is the subject of the present work.

Defects found on hydrogen-terminated samples can take the form of sub-surface or surface charge centers which can affect the operation of nearby electric field sensitive atomic devices,[24] or as non-charged surface irregularities which limit the space available for patterning. Manually locating and characterizing defects is quite labor intensive and depends on the random distribution of these defects and the cleanliness of the terminated sample. Initial attempts to automate surface defect recognition relied on fast Fourier transforms;[25] however, the characterization of defects was limited to a few of many different species. More recently, machine learning has been applied to assist in classification and analysis of surface structures using SPM,[26,27,28,29,30] but it has yet to be applied to surface features of the H-Si(100)-2x1 surface. Here, we implement an encoder-decoder type convolutional neural network (CNN)[31,32,33] to locate and classify features on the surface. By using semantic segmentation,[34,35] the neural network is trained to recognize a variety of charged and uncharged defects commonly found on the H-Si(100) surface. After implementing the model with existing patterning,[5] and probe tip forming suites,[23] full automation of the patterning process is achieved.

Crystalline silicon is tetravalent and forms a diamond lattice; each silicon atom shares 4 bonds, two above and two below the atom. At the (100) surface, two of these bonds are unsatisfied so the crystal reorganizes to a lower energy configuration. The addition of atomic hydrogen to the silicon surface during the annealing process results in the formation of one of three possible phases. The likelihood of forming such phases can be controlled by the annealing temperature at which the sample is prepared. On a silicon surface with (100) orientation, the 2x1 phase forms at ~377 ℃, the 3x1 phase at ~127 ℃, and the 1x1 phase below ~20 ℃.[36,14,37] The most commonly used for DB patterning is the 2x1 reconstruction where each surface atom pairs with a neighboring surface atom to create a dimer pair. The dimer pairs are assembled in rows which run parallel to each other across the surface. The unsatisfied bond of each silicon atom can either be terminated with hydrogen or left vacant creating a dangling bond. Although the preparation of the H-Si(100)-2x1 phase is well understood, some surface defects decorate the otherwise perfectly clean, defect-free surface. We are able to image the defects as well as clean H-Si(100) using a STM.

In order to train the CNN to recognize these surface defects, they must be labeled in a pixel-wise manner in the STM images. Our neural network is trained with seven different classes of defects. The first is regular, clean H-Si(100)-2x1 (Figure 1a). There are two types of charged defects labelled 'type 2'[38,24] (Figure 1b), the origin of which is yet to be confirmed, and 'DB' or dangling bond[19,39] (Fig. 1c). The remainder of the known surface defects are understood to exist in a neutral charge state and consist of diversely reconstructed H-Si, adatoms, and adsorbed molecules. Figure 1d shows a 'dihydride' in which two silicon atoms each bind to 2 hydrogen atoms instead of forming a dimer pair.[40] Figure 1e shows a 'step-edge' which is a drop in the surface height by one atomic layer. Dimer rows run perpendicular to the original direction above the step and the boundary of the step edge is often marked with 1x1 or 3x1 reconstruction.[41,42,43] Figures 1f-j show several different defects that either appear too infrequently, or are often found too close to each other to properly separate them for labelling (Figure 1k). These defects and any others that were unknown were assigned the label 'clustered'. A more complete analysis of these defects will be the subject of future works. Figure 1l shows the final label class, an adsorbed species, molecule, or cluster of atoms of unknown origin labelled as 'impurity'. These defects are thought to be something other than H-Si and can usually be reduced by eliminating any potential contaminants during sample preparation.

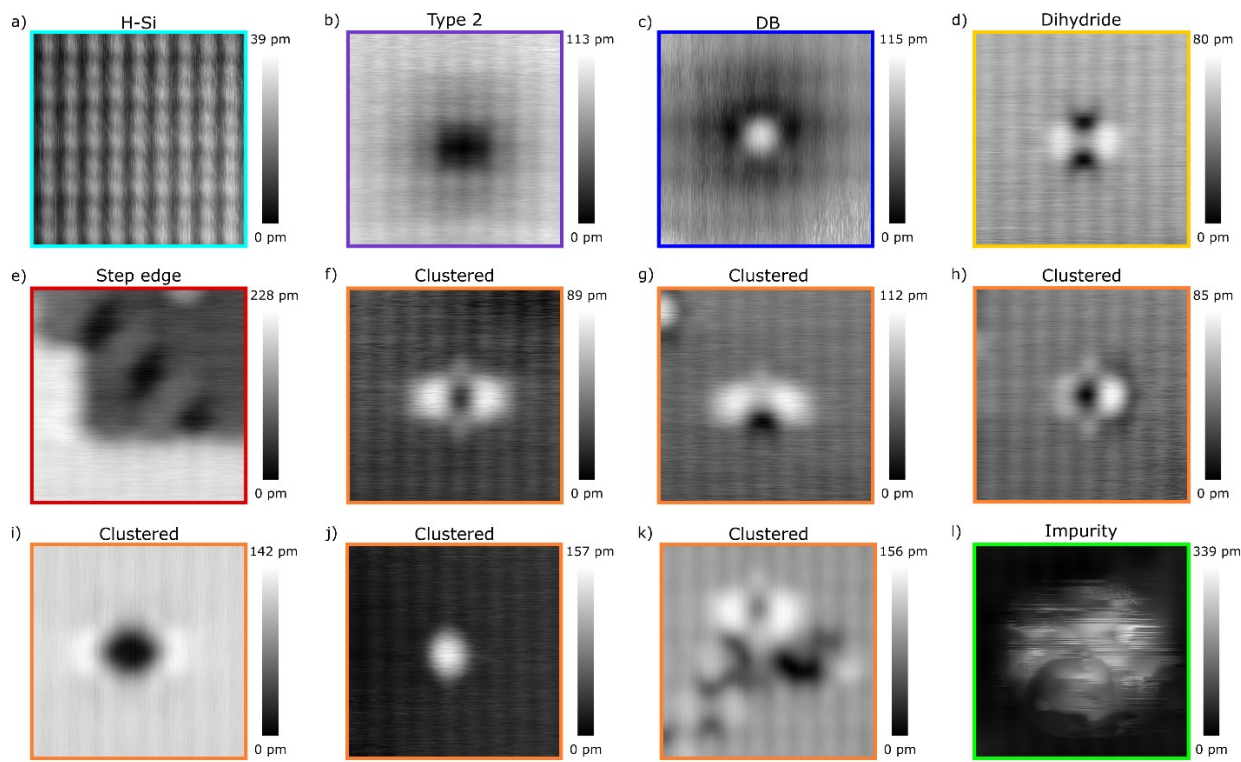

**Figure 1. Defect Classes:** Empty state images of various defects found on a typical H-Si(100) surface labelled with their appropriate class. Each image is 4 x 4 nm$^2$ taken with a STM with a sample bias of 1.4 V and a tunneling current of 50 pA. The colored borders match the assigned contour colors of Figure 3. A normalized version is shown in Figure S1.

**Methods**

All experiments and were performed using an Omicron Low Temperature STM operating at 4.5 K and ultrahigh vacuum (4 x 10[-11] Torr). Tips were electrochemically etched from polycrystalline tungsten wire and resistively heated in ultrahigh vacuum to remove surface adsorbates and oxide, and sharpened to a single atom apex using field ion microscopy.[44] *In situ* tip processing was performed by controlled tip contact with the surface.[13,45,46] Tip shaping parameters were the same as in.[23]

Samples used were highly arsenic doped (1.5 x 10[19] atoms/cm$^3$) Si(100). Samples were degassed at 600 °C overnight followed by flash annealing at 1250 °C. The samples were then terminated with hydrogen by exposing them to atomic hydrogen gas at 330 °C. It should be noted that these sample preparation guidelines were only loosely followed for all samples shown in this paper in order to ensure that a significant number of surface defects were present on the sample.

Image and data acquisition was done using a Nanonis SPM controller and software. All training data was acquired at an imaging bias of either 1.3 V or 1.4 V with a tunneling current of 50 pA. The patterning automation routine was programmed in Python and Labview using the Nanonis programming interface library.

The CNN was implemented using Keras (2.1.3) with TensorFlow backend. Data was labelled using LabelMe software.[47]

**Results and Discussion**

NEURAL NETWORK ACHRITECTURE

The architecture of the CNN was implemented to support semantic segmentation of the images. Semantic segmentation allows for both the localization and classification of objects in images. This can be used in many applications where the network must make a distinction between different objects in an image including use in self-driving cars[48,49,50] and medical image analysis.[51,52,53] In our case, a distinction is made between the pixels that make up each of the labelled defects. We trained various CNN architectures (Figure S5) and implemented the one that shows the greatest performance in defect recognition (Figure 2) (Labelled model 8 in the S.I.). An encoder-decoder type architecture is used which allows for higher order feature extraction while minimizing the number of trainable parameters.[31,54,33] Each encoder layer consists of two sets of a convolutional layer (3x3 kernel), batch normalization layer, and a 'relu' activation layer followed by a max-pooling layer (2x2 kernel). The number of convolutional filters doubles with each encoder layer starting with 32 filters reaching a maximum of 128 filters. Following the encoder layers, a series of decoder layers are applied to bring the output of the network to a size which matches the input. Each decoder layer consists of an up-sampling layer (2x2 kernel), convolutional layer (3x3 kernel), batch normalization layer, another convolution layer (3x3 kernel) and batch normalization layer followed by a relu activation layer. The final output layer consists of a convolutional layer which uses 7 filters (3x3 kernel) followed by a softmax activation which produces a one to one mapping of the surface for each of the labelled classes.

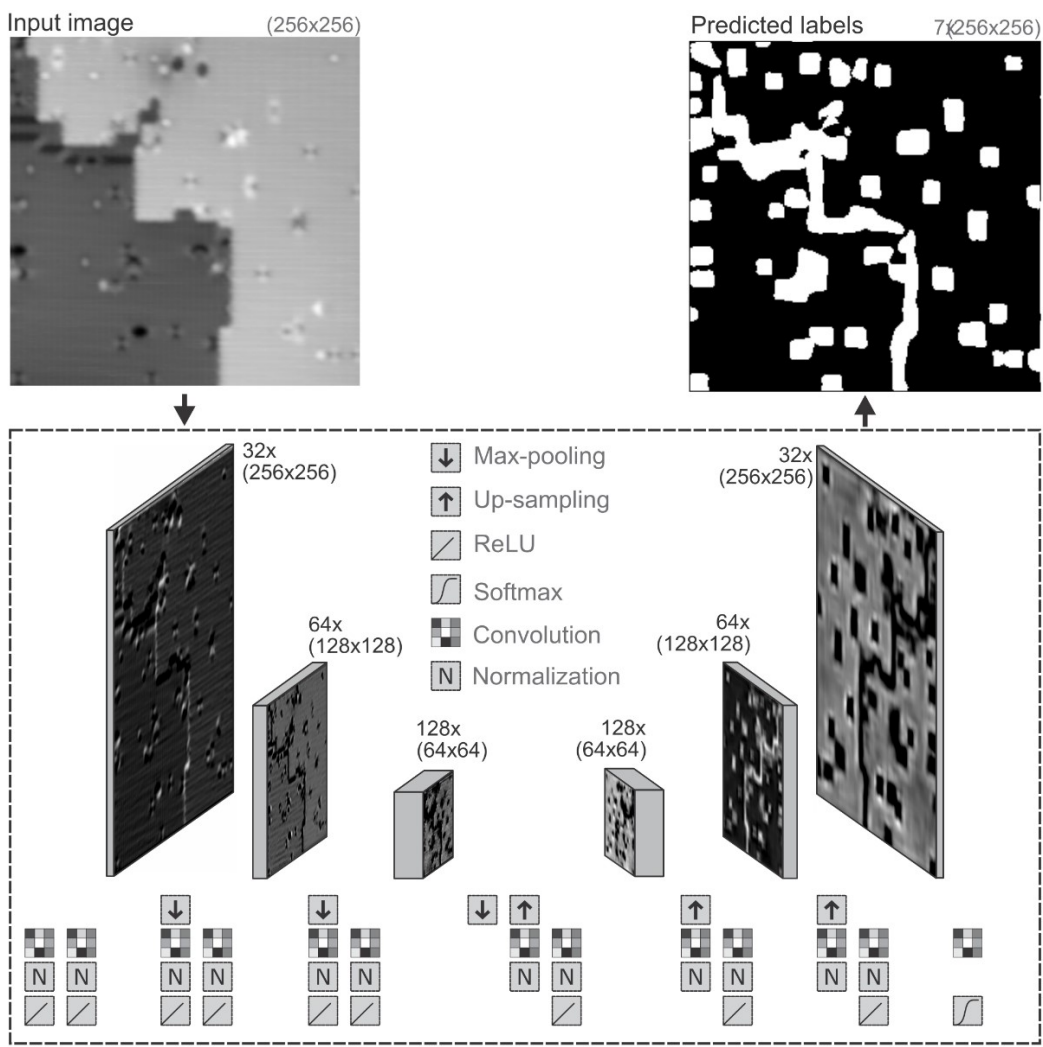

**Figure 2.** A representation of the CNN architecture used in this study. This architecture is selected by comparing the performance of different CNNs (Table S1) as well as traditional machine learning methods (We include in the SI, our attempt at using SIFT features[55] to classify surface defects). It consists of 3 convolutional encoder layers followed by 3 convolutional decoder layers. The final set of images is passed through one final convolutional layer followed by a softmax activation giving 7 separate images corresponding to each of the 7 labels. The output displayed here marks the clean H-Si in black.

DATA SET AND TRAINING

The network training data set was compiled from 28 images (100 x 100 nm$^2$ with a resolution of 1024 x 1024). Each of the 28 images were divided into 64 smaller images (128 x 128). Each of the smaller images in the training set were rotated by 90°, 180°, and 270° as well as flipped along its axis and subsequently rotated increasing our training data by a factor of 8. Images were divided into training, testing, and validating images at a ratio of ~ 2/3:1/6:1/6 respectively (corresponding to 9560:2384:2393 images). Although all images used in the training set are of the same size, the network was designed to take images of varying sizes as inputs. The only restriction to the input images is that they all have the same resolution of 1024 pixels/100 nm. This ensures that each convolutional filter will extract the same feature profiles on a variety of STM images. We utilized the Adam optimization algorithm[56] with learning rate of 0.01. Subsequent model retraining was done using a learning rate of 0.001 which very slightly improved network performance in this case. The networks were trained using a categorical cross entropy loss function. The network quality was assessed using a soft Dice loss function, to reduce the effect of the large class imbalance found in our training data.[57]

NEURAL NETWORK PERFORMANCE

A subset of the outputs of our fully trained model can be seen in Figure 3. The clean H-Si label was left out because of overlapping boundaries with the defects. More examples of the predicted label outputs including clean H-Si can be seen in Figures S2 and S3. The overall Dice score of the model is recorded at 0.86 which was calculated as a weighted average of the Dice score for each label. A full confusion matrix showing all individual Dice scores including other network architectures can be seen in Figure S6. A large portion of the 0.14 inaccuracy can be attributed to the fact that we had multiple users labeling data without a standardized defect size in place. This effect can be seen when comparing the labelled test data set with the predicted labels (Figure S4). The edges of the labelled data are straight, while the predicted label edges show a much rougher border. One would expect that if each of the defect types were traced with a constant label size, the predicted edges would better replicate the labelled edges. This effect can be seen in the confusion tables (Figure S6). Lower scores are observed for defects with a high edge-to-surface pixel ratio (type 2, DBs, dihydrides) compared to defects with a lower edge-to-surface pixel ratio (H-Si). For our purposes, this does not present an issue as the size of the defects are much larger than the variation in the predicted defect edges.

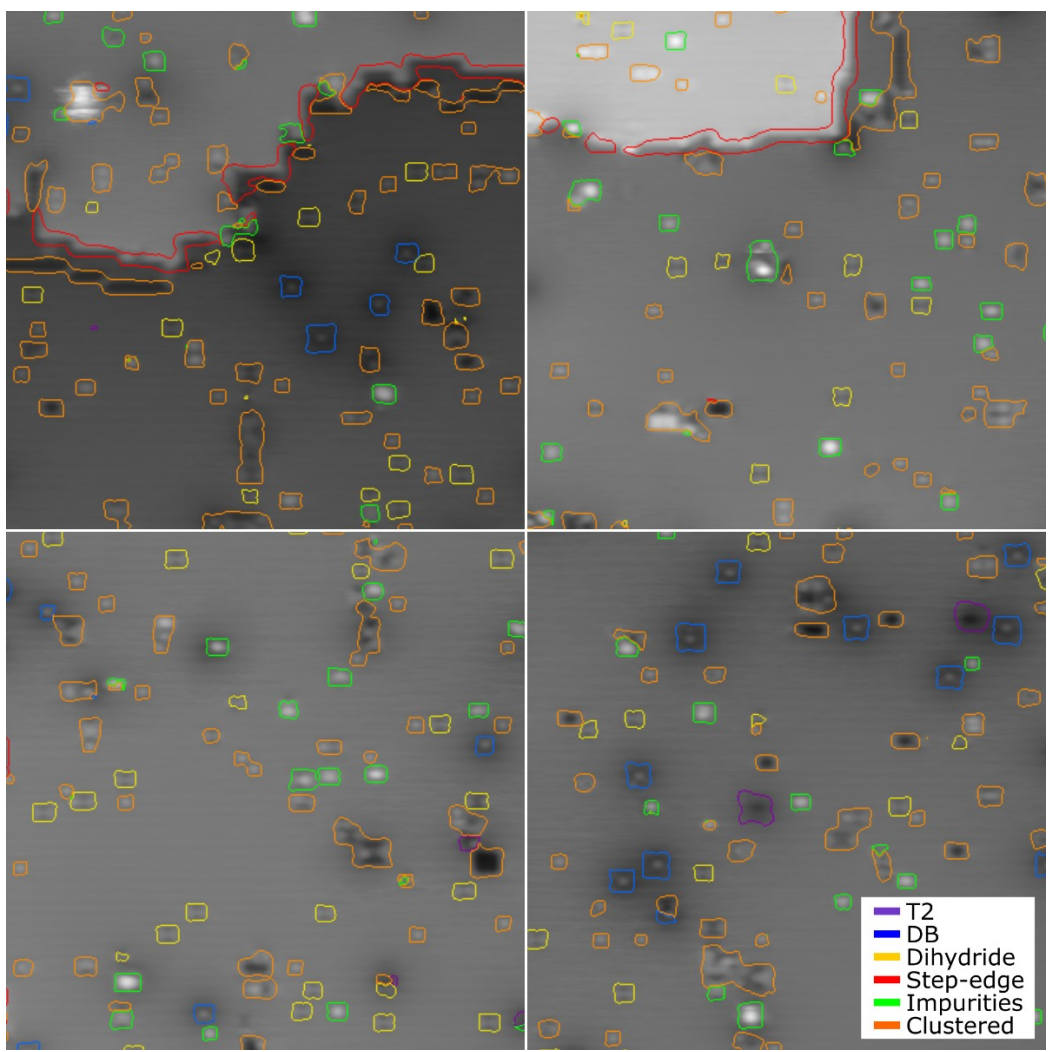

**Figure 3.** Traces of the predicted labels of the CNN and the original input images (constant current of 50 pA and 1.4 V). Each image is 40 x 40 nm$^2$

# AUGMENTATION WITH SCANNING PROBE LITHOGRAPHY

With the successful development of the neural network, it was implemented in the automation of hydrogen lithography. Figure 4 summarizes the current automation process. The user inputs a pre-designed pattern they wish to create (inset of Figure 4b), and initiates the patterning process. The SPM controller then takes a scan of the sample with a resolution matching the training data. The image is fed to the neural network and an output image containing each of the defects is returned. In order to decrease the local electrostatic interactions of local charge defects with the DB pattern,[24] an effective radius of ~5nm is applied to all type 2 defects on the surface. This increased spacing does not need to be applied to DBs since they are now routinely erasable. The same radius is applied to step-edges as well to allow space to ensure all subsequent DB patterns are made on the same step terrace. The program then identifies the area on the sample furthest from any defects that would support the pattern (white box in Figure 4a). Once found, the program begins patterning until it is complete (Figure 4b). A full flow chart of the patterning program can be seen in the SI (Figure S7) as well as additional patterning examples (Figure S9-S12). The same procedure could be applied to more complicated fabrication schemes.

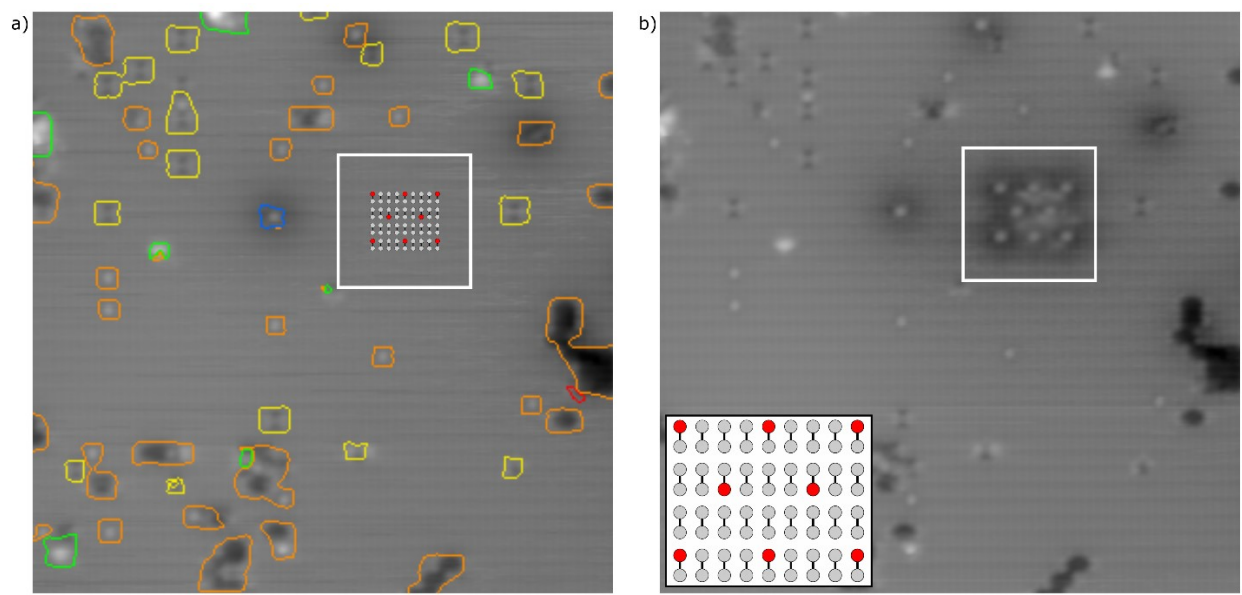

**Figure 4.** a) A trace of the defects from the CNN analysis of the scan image shown. The white square shows the area furthest from defects for the corresponding pattern. b) The resulting surface after patterning the device shown in the inset.

**Conclusion**

Continuing on the path to fully develop atomically-precise fabrication tools, we have successfully implemented a routine that can assess the quality of a sample, identify a suitable area that is free of defects, and execute a hydrogen lithography procedure. The routine is based on a CNN which uses semantic segmentation to locate and differentiate between certain charged and uncharged defects that inhibit the manufacturing process or potentially alter the operation of patterned devices. We have demonstrated the applicability of our approach by training the neural network with images of defects commonly found on the H-Si(100)-(2x1) surface. Hydrogen lithography was shown by patterning an 8 DB structure on the H-Si surface. It is envisioned that defect-free regions adequate for fabrication of functional logic or memory units comprised of roughly one hundred atoms will exist and that interconnections between such units will be custom routed so as to avoid defects. In this way, defect-free surface areas will be connected to form larger, effectively defect-free circuit blocks. In addition to avoiding defects, erasure of certain defects identified using the neural network is expected to become fully automated in future works. The techniques shown here are applicable to any type of device fabrication or lithography using any form of scanning probe microscopy as well as subsets of semiconductor device fabrication where the quality of the materials used must be assessed to optimize the fabrication process.

## Acknowledgments

We thank NSERC, AITF, and Compute Canada for financial support. We would also like to thank Mark Salomons and Martin Cloutier for their technical expertise and to Thomas Dienel and Taleana Huff for valuable discussions.

## Author Contributions

‡ M.R. and J.C. contributed equally to this work. M.R. and J. C. developed the code. J. C., M. R., K. M., M. T., and H. H. identified and labeled the defects. R.A.W. supervised the project. J.C. and M.R. wrote the manuscript with input from all authors.

## Competing Interests

The authors declare competing financial interests: A patent has been filed on this subject. Some of the authors are affiliated with Quantum Silicon Inc (QSi). QSi is seeking to commercialize atomic silicon quantum dot based technologies.

## Corresponding Author

*Correspondence can be directed to J.C. at croshaw@ualberta.ca or R.A.W. at rwolkow@ualberta.ca

**Data Availability**

Data is available upon request from the corresponding authors.

**References**

1.  Zwanenburg, F. A. *et al.* Silicon quantum electronics. *Rev. Mod. Phys.* **85**, 961–1019 (2013).

2.  Schroder, D. K. *Semiconductor Material and Device Characterization*. (John Wiley & Sons, Inc., 2005).

3.  Prati, E. & Shinada, T. Atomic scale devices: Advancements and directions. *IEEE Int. Electron Devices Meet.* 1.2.1-1.2.4 (2015).

4.  Kalff, F. E. *et al.* A kilobyte rewritable atomic memory. *Nat. Nanotechnol.* **11**, 926–929 (2016).

5.  Achal, R. *et al.* Lithography for robust and editable atomic-scale silicon devices and memories. *Nat. Commun.* **9**, 1–8 (2018).

6.  Khajetoorians, A. A., Wiebe, J., Chilian, B. & Wiesendanger, R. Realizing All-Spin-Based Logic Operations Atom by Atom. *Science (80-. ).* **332**, 1062–1064 (2011).

7.  Fuechsle, M. *et al.* A single-atom transistor. *Nat. Nanotechnol.* **7**, 242–246 (2012).

8.  Huff, T. *et al.* Binary atomic silicon logic. *Nat. Electron.* **1**, 636–643 (2018).

9.  Imre, A. *et al.* Majority Logic Gate for Magnetic Quantum-Dot Cellular Automata. *Science (80-. ).* **331**, 205–208 (2006).

10. Sotthewes, K., Geskin, V., Heimbuch, R., Kumar, A. & Zandvliet, H. J. W. Research Update: Molecular electronics: The single-molecule switch and transistor. *APL*

*Mater.* **2**, (2014).

11. Fölsch, S., Yang, J., Nacci, C. & Kanisawa, K. Atom-By-Atom Quantum State Control in Adatom Chains on a Semiconductor. *Phys. Rev. Lett.* **103**, (2009).

12. Pavliček, N., Majzik, Z., Meyer, G. & Gross, L. Tip-induced passivation of dangling bonds on hydrogenated Si(100)-2 × 1. *Appl. Phys. Lett.* **111**, (2017).

13. Huff, T. R. *et al.* Atomic White-Out: Enabling Atomic Circuitry through Mechanically Induced Bonding of Single Hydrogen Atoms to a Silicon Surface. *ACS Nano* **11**, 8636–8642 (2017).

14. Haider, M. B. *et al.* Controlled Coupling and Occupation of Silicon Atomic Quantum Dots at Room Temperature. *Phys. Rev. Lett.* **102**, 1–4 (2009).

15. Schwalb, C. H., Dürr, M. & Höfer, U. High-temperature investigation of intradimer diffusion of hydrogen on Si(001). *Phys. Rev. B* **82**, (2010).

16. Lopinski, G. P., Wayner, D. D. M. & Wolkow, R. A. Self-directed growth of molecular nanostructures on silicon. *Nature* **406**, 48–51 (2000).

17. Tong, X., Dilabio, G. A. & Wolkow, R. A. A Self-Directed Growth Process for Creating Covalently Bonded Molecular Assemblies on the H-Si(100)-3×1 Surface. *Nano Lett.* **4**, 979–983 (2004).

18. Sieval, A. B. *et al.* Monolayers of 1-Alkynes on the H-Terminated Si(100) Surface. *Langmuir* **16**, 10359–10368 (2000).

19. Lyding, J. W., Shen, T.-C., Abeln, G. C., Wang, C. & Tucker, J. R. Nanoscale patterning and selective chemistry of silicon surfaces by ultrahigh-vacuum scanning tunneling microscopy. *Nanotechnology* **7**, 128 (1996).

20. Schofield, S. R. *et al.* Atomically Precise Placement of Single Dopants in Si. *Phys. Rev. Lett.* **91**, (2003).

21. Wolkow, R. A Step Toward Making and Wiring Up Molecular-Scale Devices. *Jpn. J. Appl. Phys.* **40**, 4378–4380 (2001).

22. Tucker, J. . & Shen, T.-C. Prospects for Atomically Ordered Device Structures Based on STM Lithography. *Solid State Electron.* **42**, 1061–1067 (1998).

23. Rashidi, M. & Wolkow, R. A. Autonomous Scanning Probe Microscopy in situ Tip Conditioning through Machine Learning. *ACS Nano* **12**, 5185–5189 (2018).

24. Huff, T. Electrostatic Landscape of a Hydrogen-Terminated Silicon Surface Probed by a Moveable Quantum Dot. *ACS Nano* **13**, 10566-10575 (2019)

25. Randall, J. N., Von Ehr, J. R., Ballard, J. B., Owen, J. H. G. & Fuchs, E. Automated Scanning Tunneling Microscope image analysis of Si (100):H 2 x 1 surfaces. *Microelectron. Eng.* **98**, 214–217 (2012).

26. Modarres, M. H. *et al.* Neural Network for Nanoscience Scanning Electron Microscope Image Recognition. *Sci. Rep.* **7**, 1:12 (2017).

27. Ziatdinov, M., Maksov, A. & Kalinin, S. V. Learning surface molecular structures via machine vision. *npj Comput. Mater.* **3**, (2017).

28. Ziatdinov, M. *et al.* Deep Learning of Atomically Resolved Scanning Transmission Electron Microscopy Images: Chemical Identification and Tracking Local Transformations. *ACS Nano* **11**, 12742–12752 (2017).

29.  Gordon, O. *et al.* Scanning Probe State Recognition With Multi-Class Neural Network Ensembles. *Arxiv:* (2019).

30.  Dyck, O. *et al.* Atom-by-atom fabrication with electron beams. *Nat. Rev. Mater.* **4**, 497–507 (2019).

31.  Badrinarayanan, V., Kendall, A. & Cipolla, R. SegNet: A Deep Convolutional Encoder-Decoder Architecture for Image Segmentation. *IEEE Trans. Pattern Anal. Mach. Intell.* **39**, 2481–2495 (2017).

32.  Long, J., Shelhamer, E. & Darrell, T. Fully convolutional networks for semantic segmentation. in *2015 IEEE Conference on Computer Vision and Pattern Recognition (CVPR)* 3431–3440 (IEEE, 2015).

33.  Noh, H., Hong, S. & Han, B. Learning Deconvolution Network for Semantic Segmentation. in *2015 IEEE International Conference on Computer Vision (ICCV)* 1520–1528 (IEEE, 2015).

34.  Carreira, J., Caseiro, R., Batista, J. & Sminchisescu, C. Semantic Segmentation with Second-Order Pooling. in *Conference on Computer Vision and Pattern Recognition (CVPR)* 430–443 (2012).

35.  Girshick, R., Donahue, J., Darrell, T. & Malik, J. Rich Feature Hierarchies for Accurate Object Detection and Semantic Segmentation. in *2014 IEEE Conference on Computer Vision and Pattern Recognition* 580–587 (IEEE, 2014).

36.  Boland, J. J. J. Structure of the H-Saturated Si(100) Surface. *Phys. Rev. Lett.* **65**, 3325–3328 (1990).

37.  Chabal, Y. J. & Raghavachari, K. New Ordered Structure for the H-Saturated

Si(100) Surface: The (3 x 1) Phase. *Phys. Rev. Lett.* **54**, 1055–1058 (1985).

38.  Sinthiptharakoon, K. *et al.* Investigating individual arsenic dopant atoms in silicon using low-temperature scanning tunnelling microscopy. *J. Phys. Condens. Matter J. Phys. Condens. Matter* **26**, 1–8 (2014).

39.  Schofield, S. R. *et al.* Quantum engineering at the silicon surface using dangling bonds. *Nat. Commun.* **4**, 1–7 (2013).

40.  Bellec, A., Riedel, D., Dujardin, G., Rompotis, N. & Kantorovich, L. N. Dihydride dimer structures on the Si(100):H surface studied by low-temperature scanning tunneling microscopy. *Phys. Rev. B* **78**, (2008).

41.  Suwa, Y. *et al.* Formation of dihydride chains on H-terminated   Si ( 100 ) − 2 × 1 surfaces: Scanning tunneling microscopy and first-principles calculations. *Phys. Rev. B* **74**, 205308 (2006).

42.  Fujimori, M., Heike, S., Suwa, Y. & Hashizume, T. Initial-Stage Dihydride Formation on Si(100)-2x1-H Surface. *Japan J. Appl. Phys.* **42**, 1387–1390 (2003).

43.  Pearson, C., Borovsky, B., Krueger, M., Curtis, R. & Ganz, E. Si(001) Step Dynamics. *Phys. Rev. Lett.* **74**, 2710–2713 (1995).

44.  Rezeq, M., Pitters, J. & Wolkow, R. Tungsten nanotip fabrication by spatially controlled field-assisted reaction with nitrogen. *J. Chem. Phys.* **124**, (2006).

45.  Jarvis, S., Sweetman, A., Bamidele, J., Kantorovich, L. & Moriarty, P. Role of orbital overlap in atomic manipulation. *Phys. Rev. B* **85**, (2012).

46.  Labidi, H. *et al.* Indications of chemical bond contrast in AFM images of a hydrogen-

terminated silicon surface. *Nat. Commun.* **8**, (2017).

47.  Russel, B. C., Torralba, A., Murphy, K. P. & Freeman, W. T. LabelMe: a database and web-based tool for image annotation. *Int. J. Comput. Vis.* **77**, 157–173 (2008).

48.  Ros, G. *et al.* Vision-based Offline-Online Perception Paradigm for Autonomous Driving. in *IEEE Winter Conference on Applications of Computer Vision* (2015).

49.  Ros, G., Sellart, L., Materzynska, J., Vazquez, D. & Lopez, A. M. The SYNTHIA Dataset: A Large Collection of Synthetic Images for Semantic Segmentation of Urban Scenes. in *Conference on Computer Vision and Pattern Recognition (CVPR)* (2016).

50.  Teichmann, M., Weber, M., Zollner, M., Cipolla, R. & Urtasun, R. MultiNet: Real-time Joint Semantic Reasoning for Autonomous Driving. in *2018 IEEE Intelligent Vehicles Symposium (IV)* 1013–1020 (IEEE, 2018).

51.  Milletari, F., Navab, N. & Ahmadi, S.-A. V-Net: Fully Convolutional Neural Networks for Volumetric Medical Image Segmentation. in *2016 Fourth International Conference on 3D Vision (3DV)* 565–571 (IEEE, 2016).

52.  Litjens, G. *et al.* A survey on deep learning in medical image analysis. *Med. Image Anal.* **42**, 60–88 (2017).

53.  Gibson, E. *et al.* NiftyNet: a deep-learning platform for medical imaging. *Comput. Methods Programs Biomed.* **158**, 113–122 (2018).

54.  Ronneberger, O., Fischer, P. & Brox, T. U-Net: Convolutional Networks for Biomedical Image Segmentation. Preprint at https://arxiv.org/abs/1505.04597 (2015). (2015).

55. Lowe, D. G. Distinctive Image Features from Scale-Invariant Keypoints. *Int. J. Comput. Vis.* **60**, 91–110 (2004).

56. Kingma, D. P. & Ba, J. Adam: A Method for Stochastic Optimization. *Int. Conf. Mach. Learn.* (2014).

57. Sudre, C. H., Li, W., Vercauteren, T., Ourselin, S. & Cardoso, M. J. Generalised Dice overlap as a deep learning loss function for highly unbalanced segmentations. Preprint at https://arxiv.org/abs/1707.03237 (2017).

# DEEP LEARNING-GUIDED SURFACE CHARACTERIZATION FOR AUTONOMOUS HYDROGEN LITHOGRAPHY:

# SUPPORTING INFORMATION

*Mohammad Rashidi[‡1], Jeremiah Croshaw[‡*1], Kieran Mastel[1], Marcus Tamura[1], Hedieh Hosseinzadeh[3], and Robert A. Wolkow[*1,2,3]*

[1]Department of Physics, University of Alberta, Edmonton, Alberta, T6G 2J1

[2]Nanotechnology Research Centre, National Research Council Canada, Edmonton, Alberta, T6G 2M9, Canada

[3]Quantum Silicon, Inc., Edmonton, Alberta, T6G 2M9, Canada

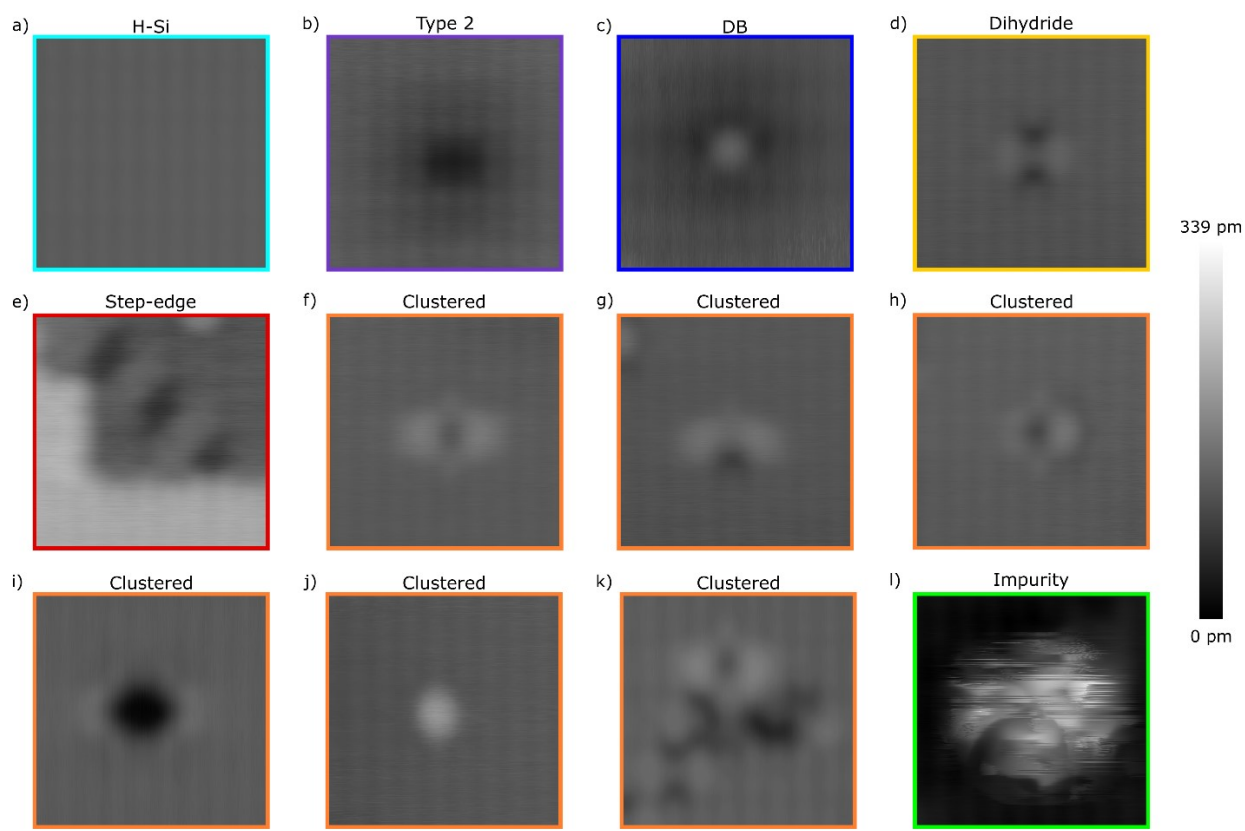

**Figure S1**. Normalized images of the defects found in figure 1. a)-k) have been normalized so that the H-Si surrounding the defect is at the same relative height. l) has been left unmodified. All images now use the same height scale.

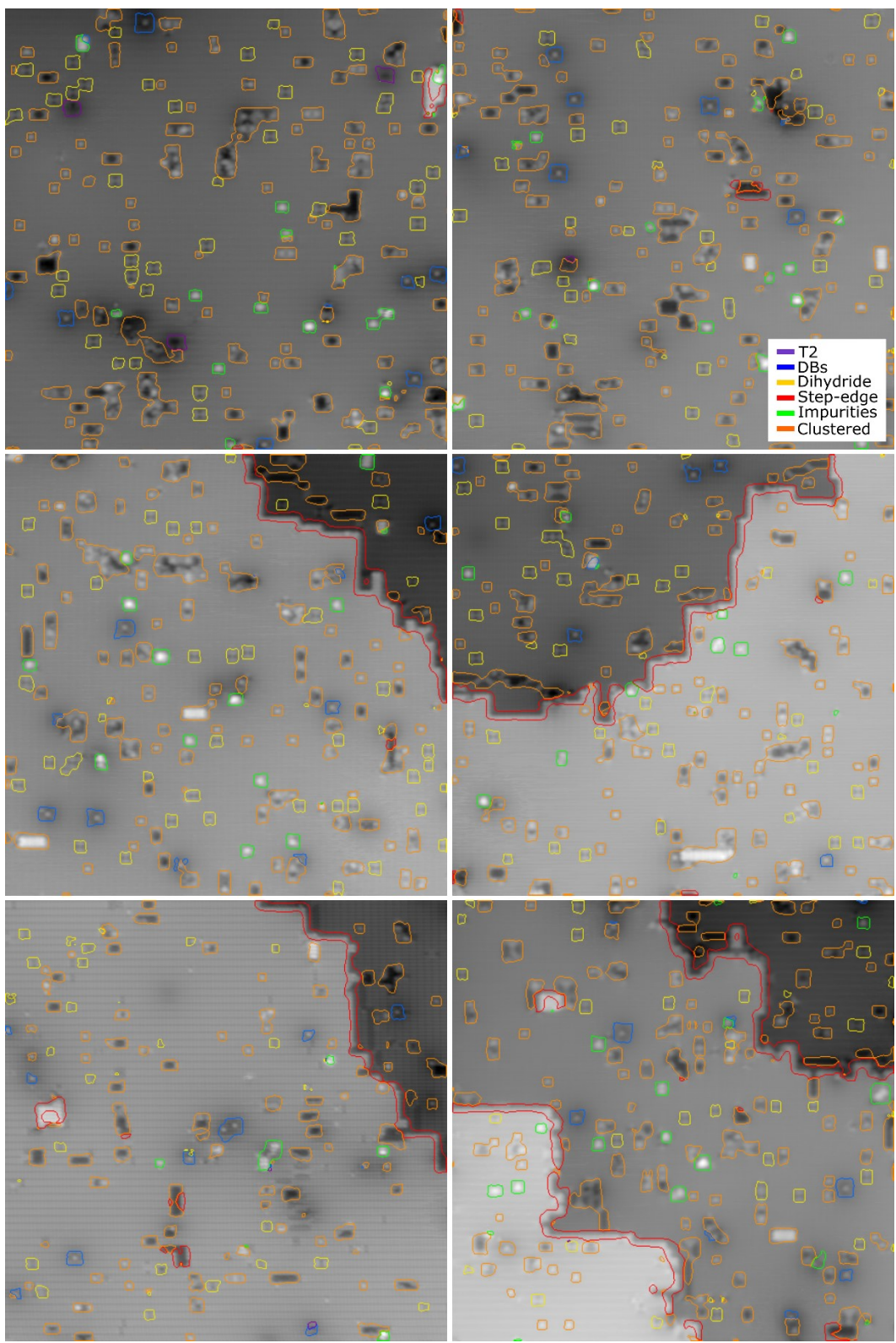

**Figure S2.** Surface scans and six of the labeled outputs from the neural network. Each image is 50 x 50 nm² and was taken at a sample bias of 1.3 V and tunneling current of 50 pA. The surface shown in the bottom left was imaged using a rare tip contrast that the network was not very familiar with. A decrease in the accuracy of defect segmentation compared to the other five images can be seen.

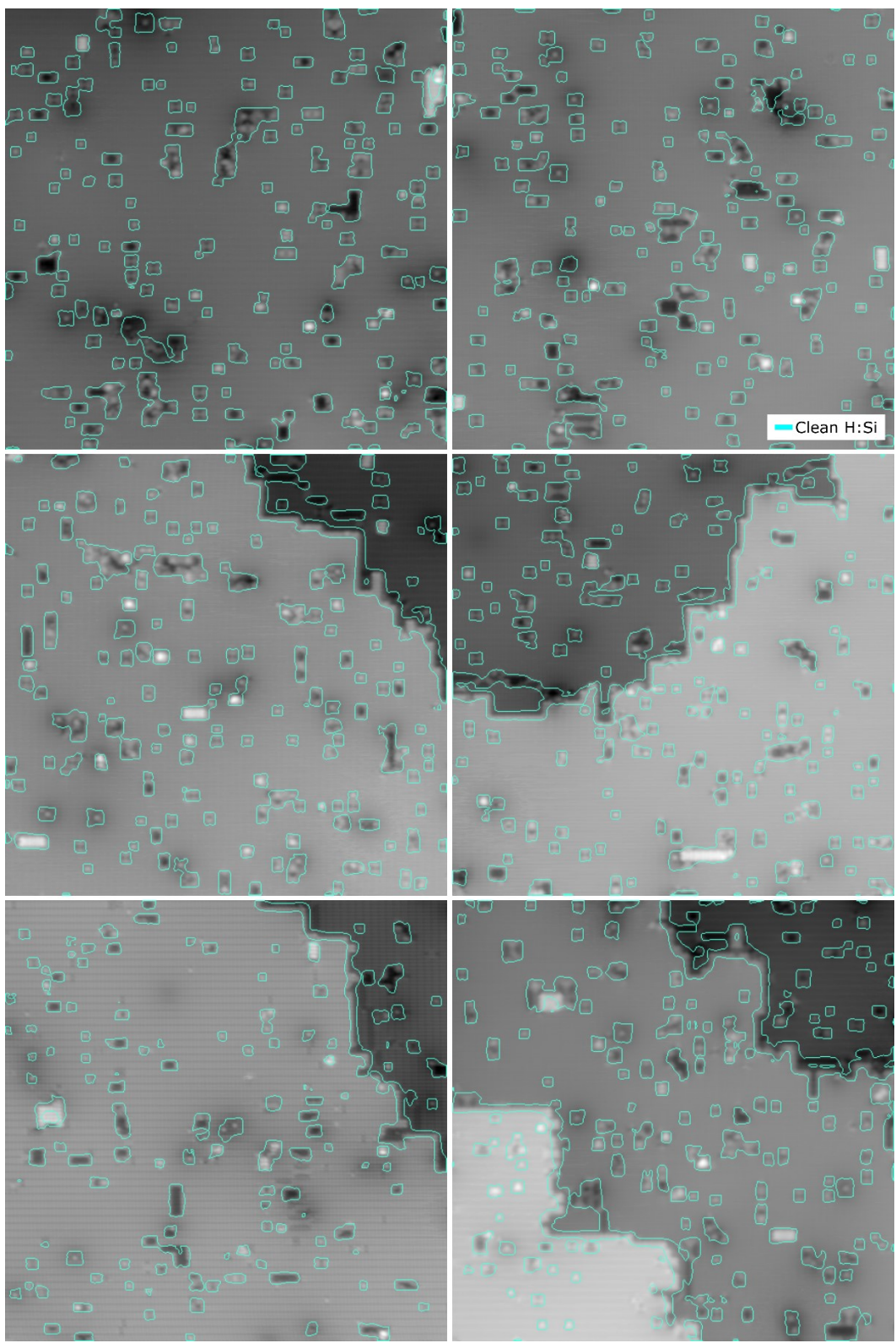

**Figure S3.** Surface scans and the H-Si output from the neural network. Each image is 50 x 50 nm² and was taken at a sample bias of 1.3 V and tunneling current of 50 pA. The surface shown in the bottom left was imaged using a rare tip contrast that the network was not very familiar with. A decrease in the accuracy of defect segmentation compared to the other five images can be seen.

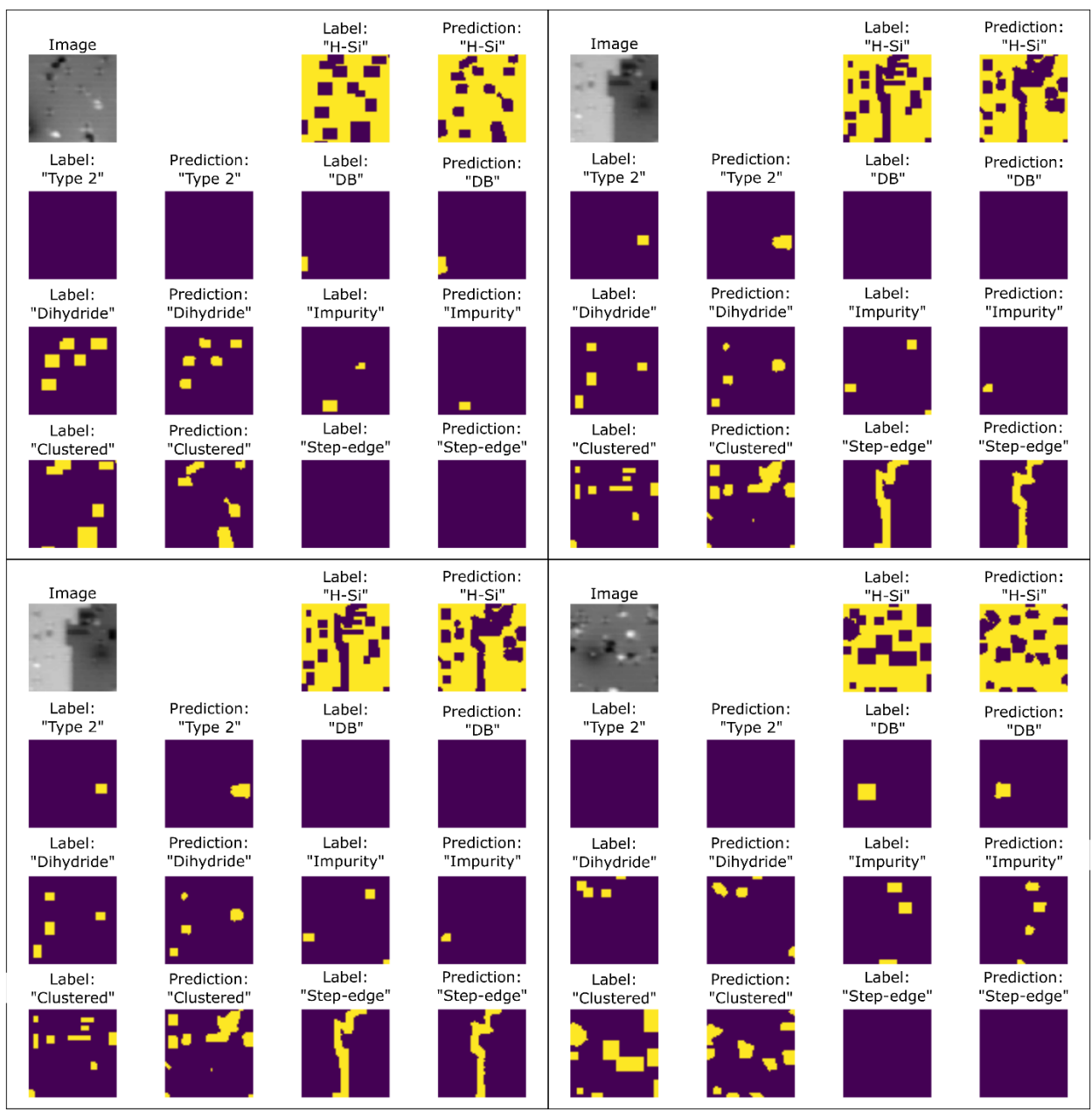

**Figure S4**: Heat maps from a random selection of the test data set. True pixels are yellow while false pixels are purple. Each image is 12.5 x 12.5 nm$^2$.

## Neural Network Architecture

A variety of neural network architectures were trained to find a configuration which gave the best Dice score of the Test and Validation data. Figure S5 shows the architectures of 10 of these models. These 10 models were chosen from the many more that were trained to highlight their superior performance or unique architecture. Models 1 through 5 where designed based on the U-net[1] model. Comparing the dice scores found in Figure S6 or Table S1, it can be seen that the implementation of skip layers did not increase the dice scores of the networks. Model 6 was based off the network presented in[2]. Models 7-9 were modifications of this network style with the addition of convolutional filters. Model 10 was a modification of model 1 without any downsampling layers. The removal of these layers made training the network much more time consuming which necessitated the removal of four convolutional layers. The two networks with the best Dice score were models 8 and 9 which used double convolution style encoder/decoder layers. Model 9 had the advantage of a quicker image processing time, however given that the current time scale to acquire an image is roughly 10 minutes, the additional 5 seconds saved will not significantly decrease the fabrication time. As a result, model 8 was chosen due to the slightly better performance at classifying T2s, DBs, dihydrides, impurities, and clustered defects as shown in the confusion tables of figure S6. Interestingly, the dice score for each defect remained relatively constant for each defect type lending to the discussion in the main text suggesting that the edge-to-surface ratio of the defects in combination with the large class imbalance place an upper limit on the dice scores achievable with this data set.

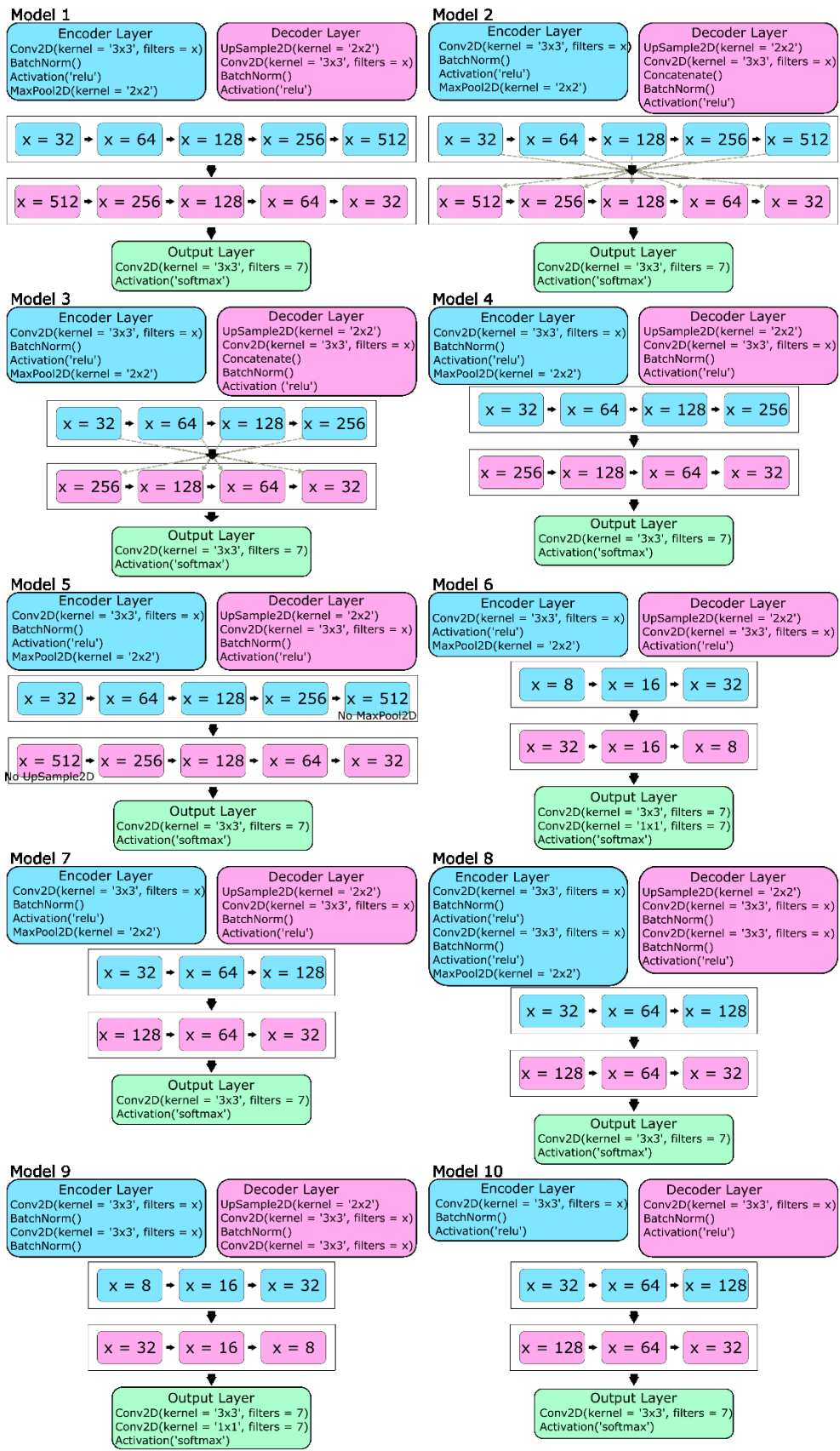

Figure S5. Architectures of each of the 10 model architectures tested. Each model shows the variation in the encoder, decoder, and output layers between each model.

**Table S1.** Parameters of each of the tested models highlighting key features of their architecture as well as classification performance.

| Model # | Convolution Layers | Pooling Layers (one direction) | Trainable Parameters | Dice Score (Test Data) | Dice Score (Val Data) | Load and Analyze Time (s) |
|---|---|---|---|---|---|---|
| 1 | 11 | 5 | 5,500,999 | 0.860 | 0.856 | 12.1 |
| 2 | 11 | 5 | 7,071,719 | 0.843 | 0.853 | 15.5 |
| 3 | 9 | 4 | 1,759,207 | 0.865 | 0.860 | 12.6 |
| 4 | 9 | 4 | 1,369,159 | 0.858 | 0.854 | 10.7 |
| 5 | 11 | 4 | 5,500,999 | 0.828 | 0.826 | 12.6 |
| 6 | 8 | 3 | 21,487 | 0.860 | 0.855 | 6.0 |
| 7 | 7 | 3 | 335,431 | 0.867 | 0.863 | 9.5 |
| 8 | 13 | 3 | 735,175 | 0.868 | 0.862 | 14.8 |
| 9 | 14 | 3 | 45,639 | 0.868 | 0.864 | 9.7 |
| 10 | 7 | 0 | 335,431 | 0.857 | 0.852 | 16.5 |

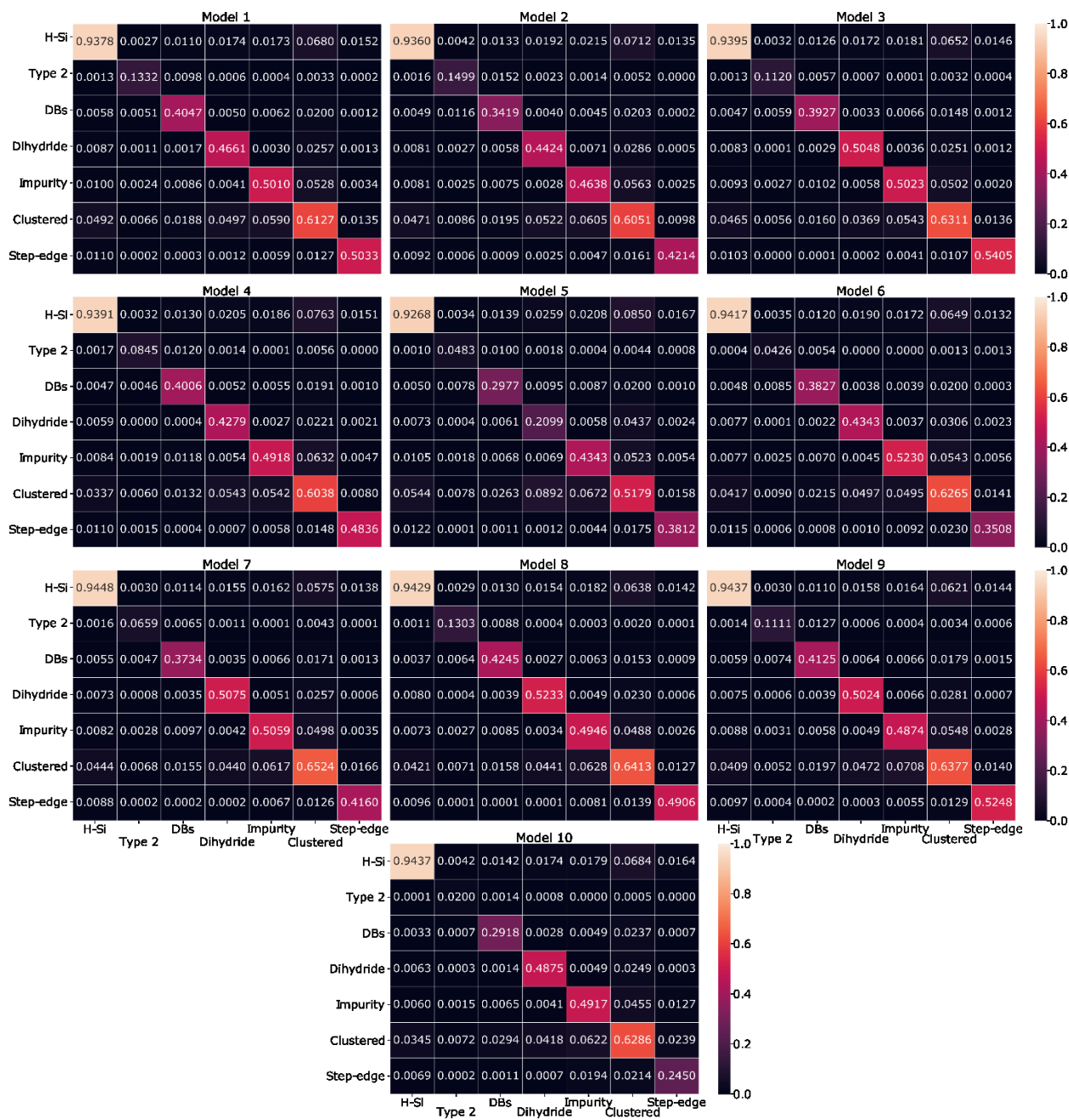

**Figure S6.** Confusion matrices showing labeling accuracy of the Test dataset for each of the 10 models. Each entry represents the Dice score for the actual label in the column being identified as the predicted label along the row.

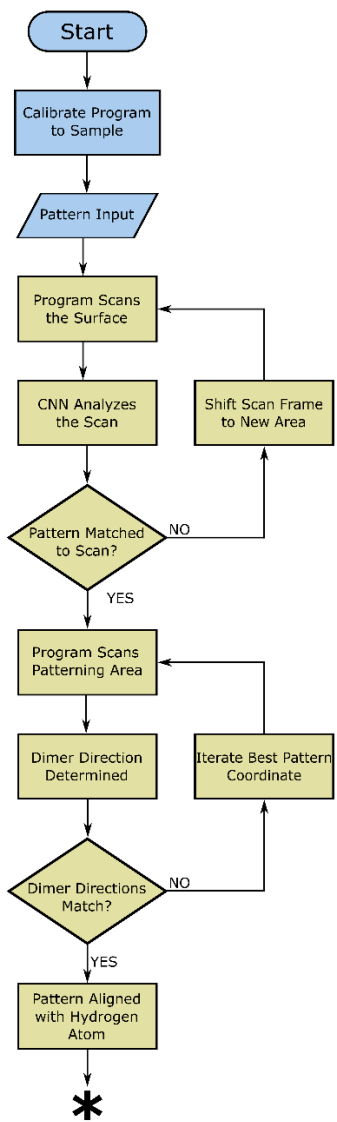
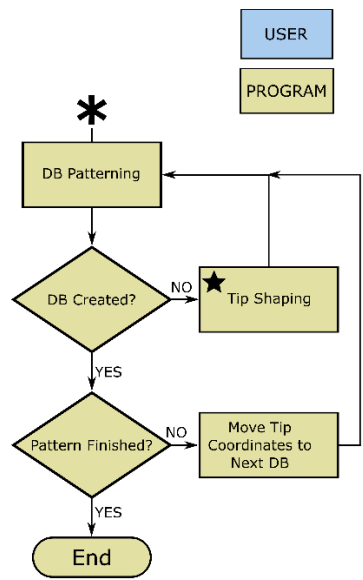
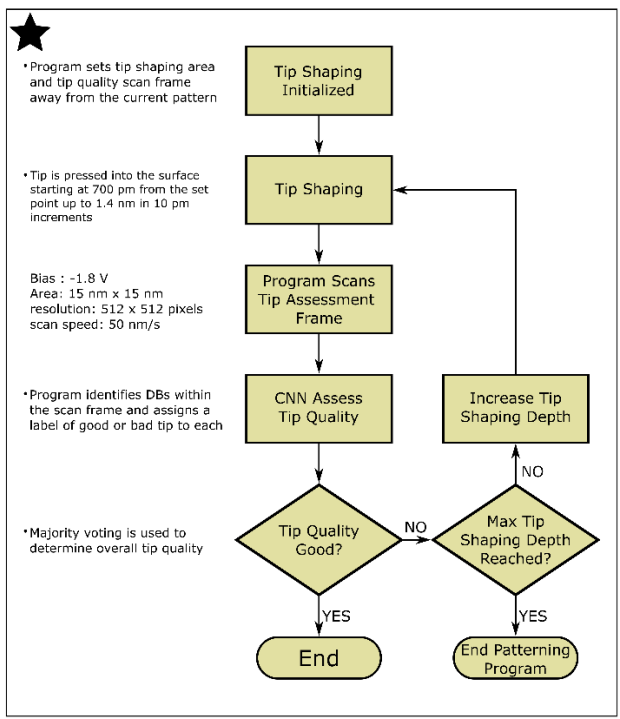

**Figure S7.** Decision tree summary followed in the lithography automation suite.

**Scale Invariant Feature Transform Classification**

In addition to neural networks, an attempt to classify surface defects using scale invariant feature transform (SIFT) descriptors[3] was also made. Training data was created from the already labelled data by taking 20 x 20 pixel cutouts of each the 6 defect types see in Figure S8a. Valid keypoints and descriptors for each defect were calculated using the SIFT library from the OpenCV package[4]. Only 64% of the defects returned valid keypoints and descriptors. Visual inspection of the images which failed to return valid SIFT keypoints shows that these defects were taken from larger images with multiple step edges resulting in very little contrast in the defect image. After splitting the defects into training and testing data (0.67:0.33), the descriptors were clustered using a mini batch k-means method using 500 clusters. From there, the frequency of descriptors in each image belonging to these clusters where counted. These frequencies were normalized using a standard scaling method and were used as features to train a random forest classifier. An overall accuracy of 68% is achieved on the test set with the full confusion table presented in Figure S8b. Even if this accuracy could be improved to a level comparable with the neural network approach, in order to achieve a similar resolution in defect segmentation, a sliding window approach is needed. Unfortunately, the SIFT algorithm available through OpenCV is a relatively slow process. Using a sliding window approach with a window size of 20 x 20 pixels, it takes roughly 2 hours to analyze a 50 x 50 nm image compared to the 15 seconds needed using neural networks with the same computational power.

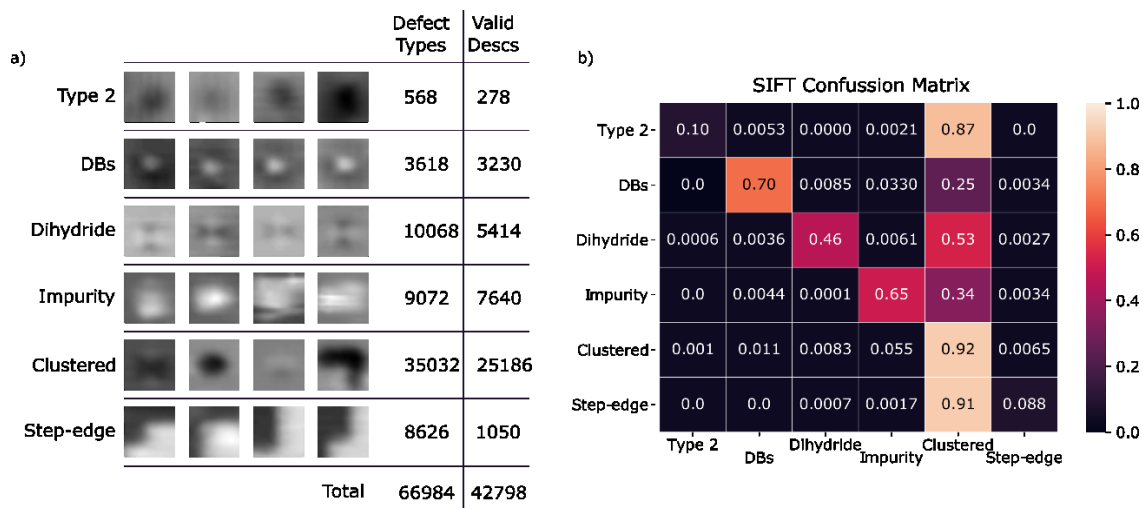

a)

| | | | | | Defect Types | Valid Descs |
|---|---|---|---|---|---|---|
| Type 2 | | | | | 568 | 278 |
| DBs | | | | | 3618 | 3230 |
| Dihydride | | | | | 10068 | 5414 |
| Impurity | | | | | 9072 | 7640 |
| Clustered | | | | | 35032 | 25186 |
| Step-edge | | | | | 8626 | 1050 |
| | | | | Total | 66984 | 42798 |

b)

**SIFT Confussion Matrix**

| | Type 2 | DBs | Dihydride | Impurity | Clustered | Step-edge |
|---|---|---|---|---|---|---|
| Type 2 | 0.10 | 0.0053 | 0.0000 | 0.0021 | 0.87 | 0.0 |
| DBs | 0.0 | 0.70 | 0.0085 | 0.0330 | 0.25 | 0.0034 |
| Dihydride | 0.0006 | 0.0036 | 0.46 | 0.0061 | 0.53 | 0.0027 |
| Impurity | 0.0 | 0.0044 | 0.0001 | 0.65 | 0.34 | 0.0034 |
| Clustered | 0.001 | 0.011 | 0.0083 | 0.055 | 0.92 | 0.0065 |
| Step-edge | 0.0 | 0.0 | 0.0007 | 0.0017 | 0.91 | 0.088 |

**Figure S8**  a) A sub-sampling of the images used to create the training data set for SIFT classification. The number of images used for each defect is also listed as well as the number of images for each defect which produced a valid descriptor using sift classification. b) The Confusion Matrix of the defect classification using the SIFT descriptors and the random forest classifier.

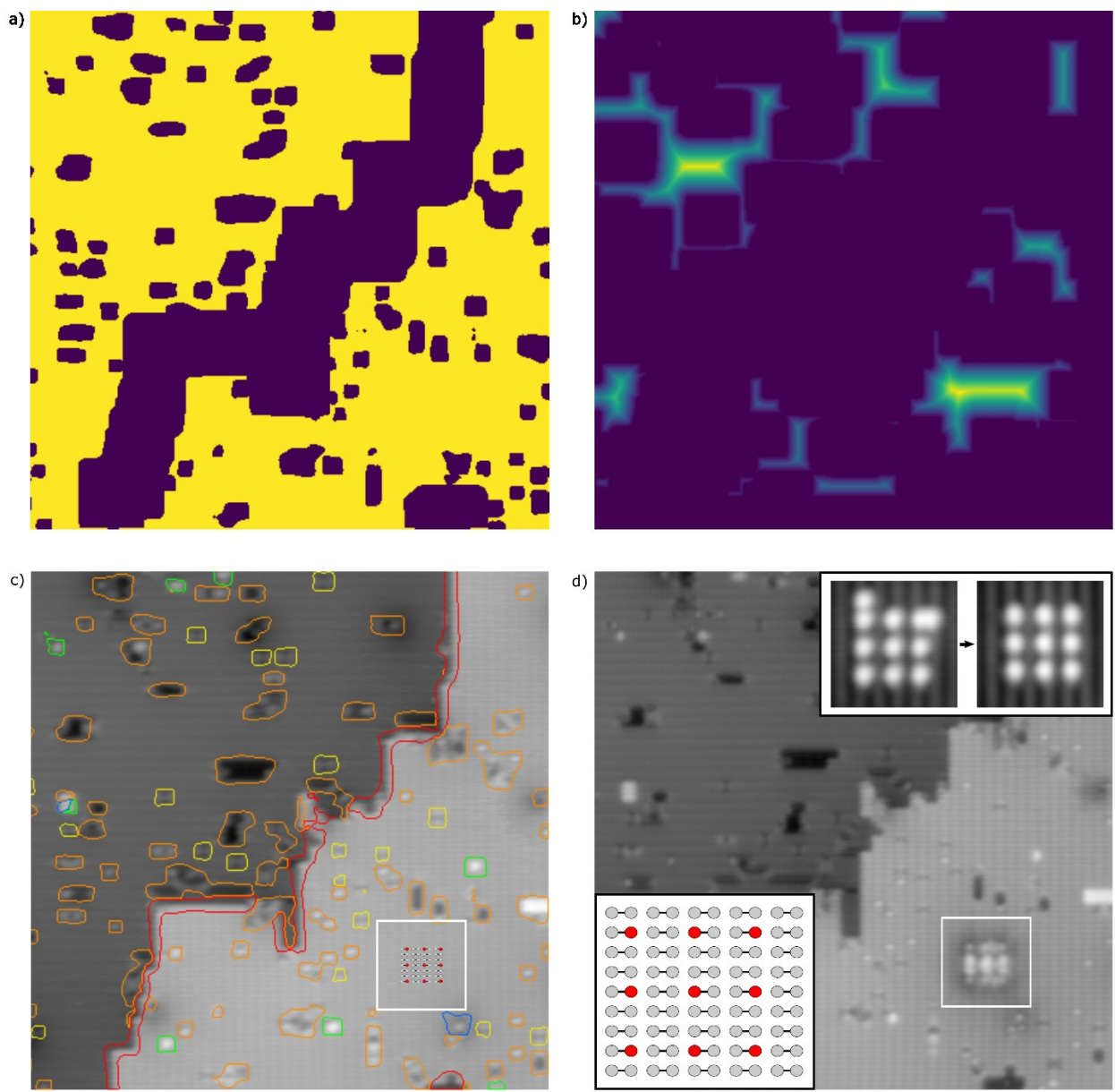

Figure S9: An outline of the fabrication procedure. a) A trace of the surface defects (purple) with the added effective thickness surrounding the step-edge present in this image. b) A mapping of the best spot to place the pattern. The best fits are in yellow with the top left corner of the pattern corresponding to the brightest spots in the image. c) Alignment of the pattern with the surface showing the segmentation of defects as returned by the network. d) The resulting surface after hydrogen lithography. The inset in the upper right shows the initial pattern with a minor mistake and the manually corrected pattern. The images are 40 x 40 nm$^2$ and acquired at an imaging bias of 1.3 V and a constant current of 50 pA.

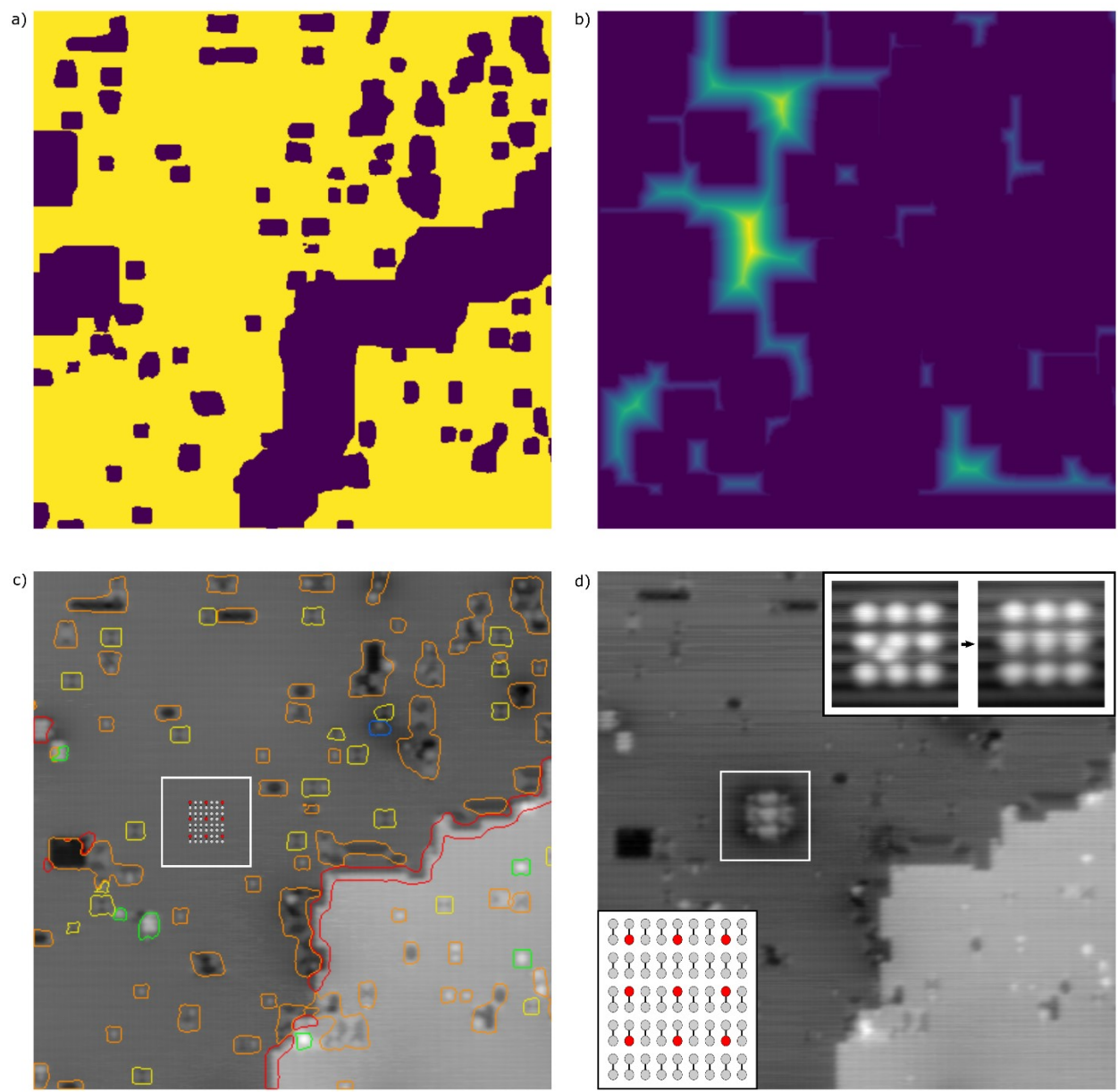

Figure S10: An outline of the fabrication procedure. a) A trace of the surface defects (purple) with the added effective thickness surrounding the step-edge present in this image. b) A mapping of the best spot to place the pattern. The best fits are in yellow with the top left corner of the pattern corresponding to the brightest spots in the image. c) Alignment of the pattern with the surface showing the segmentation of defects as returned by the network. d) The resulting surface after hydrogen lithography. The inset in the upper right shows the initial pattern with a minor mistake and the manually corrected pattern. The images are 40 x 40 nm$^2$ and acquired at an imaging bias of 1.3 V and a constant current of 50 pA.

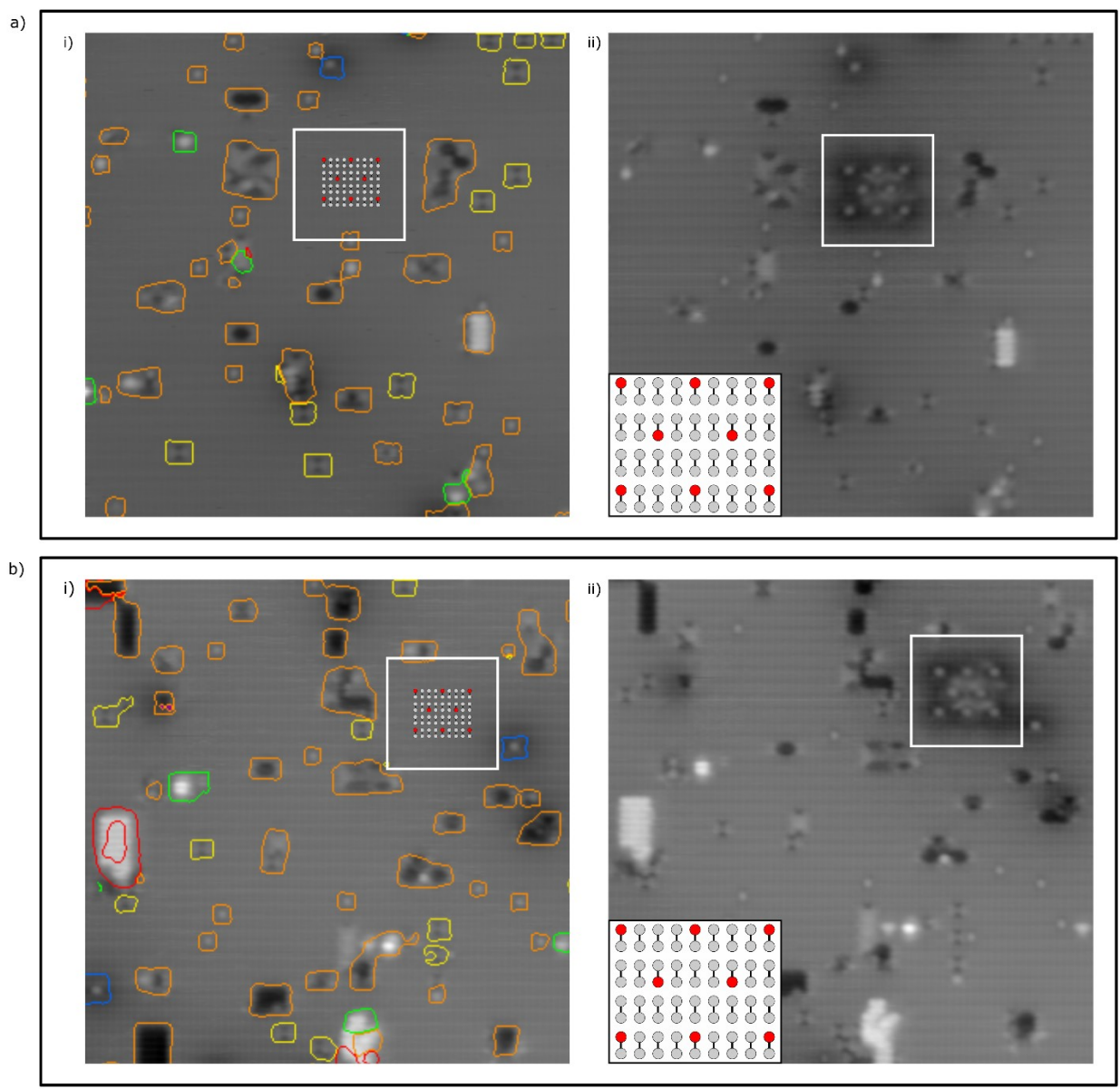

**Figure S11.** a) A trace of the defects from the CNN analysis of the scan image shown. The white square shows the area furthest from defects for the corresponding pattern. b) The resulting surface after patterning the device shown in the inset. The images are 30 x 30 nm$^2$ and acquired using a sample bias of 1.3 V and a setpoint current of 50 pA.

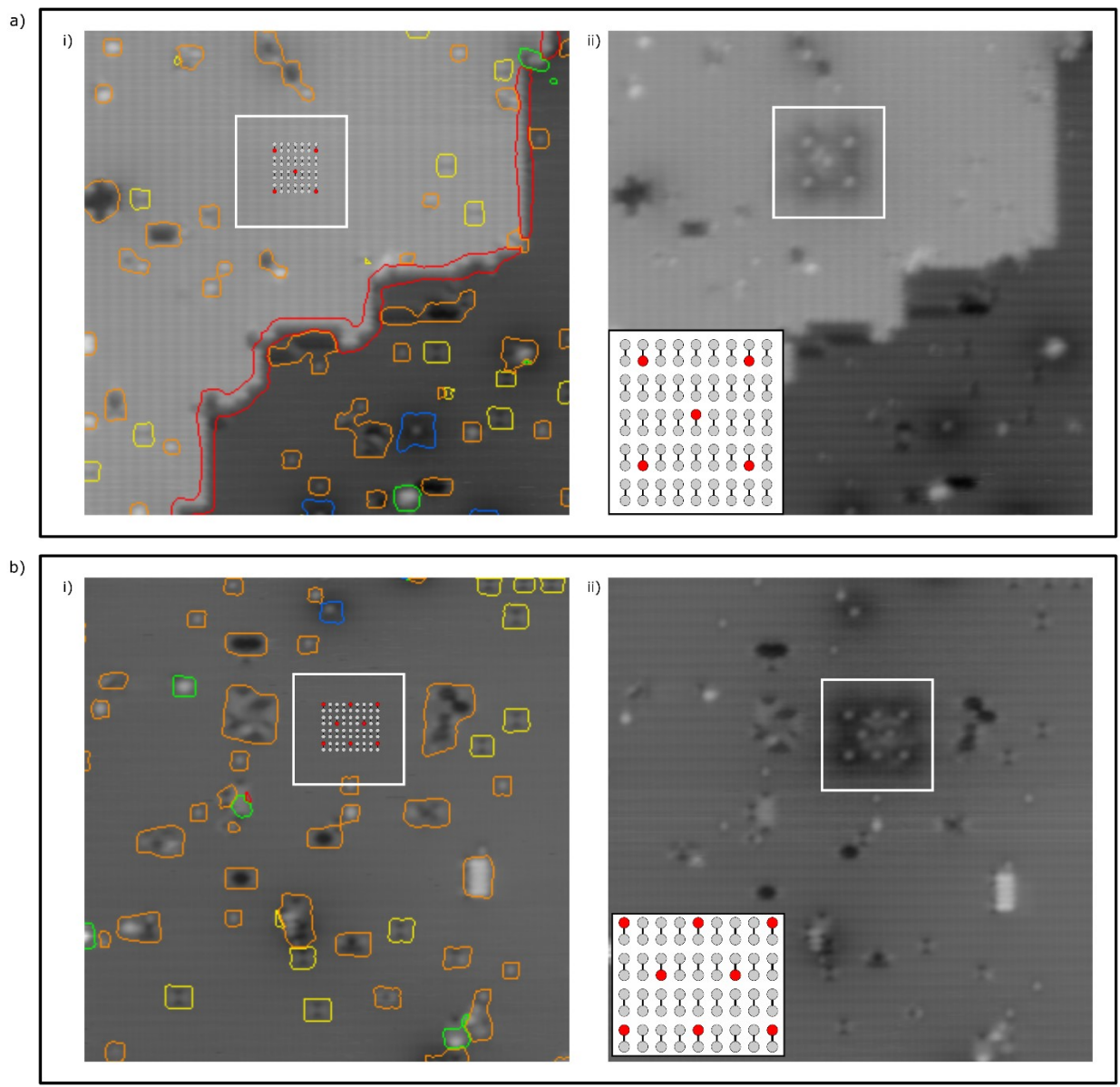

**Figure S12.** a) A trace of the defects from the CNN analysis of the scan image shown. The white square shows the area furthest from defects for the corresponding pattern. b) The resulting surface after patterning the device shown in the inset. The images are 30 x 30 nm$^2$ and acquired using a sample bias of 1.3 V and a setpoint current of 50 pA.

## References

1. Ronneberger, O., Fischer, P. & Brox, T. U-Net: Convolutional Networks for Biomedical Image Segmentation. Preprint at https://arxiv.org/abs/1505.04597 (2015).

2. Ziatdinov, M. *et al.* Deep Learning of Atomically Resolved Scanning Transmission Electron Microscopy Images: Chemical Identification and Tracking Local Transformations. *ACS Nano* **11**, 12742–12752 (2017).

3. Lowe, D. G. Distinctive Image Features from Scale-Invariant Keypoints. *Int. J. Comput. Vis.* **60**, 91–110 (2004).

4. Bradski, G. The OpenCV Library. *J. Softw. Tools* (2000).